\documentclass[a4paper, reprint, amsfonts, aps, prl, amssymb, nobibnotes, twoside, balancelastpage, amsmath] {revtex4-1}
\usepackage[utf8]{inputenc}
\usepackage[english]{babel}
\usepackage[T1]{fontenc}
\usepackage{textcomp}
\usepackage{graphicx}
\usepackage{multirow}
\usepackage[mediumspace,mediumqspace,squaren,binary]{SIunits}
\usepackage{color}
\usepackage{float}

\begin{document}                                         
\title{Selective Area Growth Rates of III-V Nanowires}

\author{Martin Espiñeira Cachaza\textsuperscript{1,2,*}}
\affiliation{\textsuperscript{1}Microsoft Quantum Materials Lab Copenhagen, 2800 Lyngby, Denmark}
\affiliation{\textsuperscript{2}Center for Quantum Devices, Niels Bohr Institute, University of Copenhagen, 2100 Copenhagen, Denmark \\ 
\textsuperscript{*}These authors contributed equally \\ 
\textsuperscript \textdagger Corresponding author: krogstrup@nbi.dk}

\author{Anna Wulff Christensen\textsuperscript{1,2,*}}
\affiliation{\textsuperscript{1}Microsoft Quantum Materials Lab Copenhagen, 2800 Lyngby, Denmark}
\affiliation{\textsuperscript{2}Center for Quantum Devices, Niels Bohr Institute, University of Copenhagen, 2100 Copenhagen, Denmark \\ 
\textsuperscript{*}These authors contributed equally \\ 
\textsuperscript \textdagger Corresponding author: krogstrup@nbi.dk}

\author{Daria Beznasyuk\textsuperscript{1,2}}
\affiliation{\textsuperscript{1}Microsoft Quantum Materials Lab Copenhagen, 2800 Lyngby, Denmark}
\affiliation{\textsuperscript{2}Center for Quantum Devices, Niels Bohr Institute, University of Copenhagen, 2100 Copenhagen, Denmark \\ 
\textsuperscript{*}These authors contributed equally \\ 
\textsuperscript \textdagger Corresponding author: krogstrup@nbi.dk}

\author{Tobias S\ae rkj\ae r\textsuperscript{1,2}}
\affiliation{\textsuperscript{1}Microsoft Quantum Materials Lab Copenhagen, 2800 Lyngby, Denmark}
\affiliation{\textsuperscript{2}Center for Quantum Devices, Niels Bohr Institute, University of Copenhagen, 2100 Copenhagen, Denmark \\ 
\textsuperscript{*}These authors contributed equally \\ 
\textsuperscript \textdagger Corresponding author: krogstrup@nbi.dk}

\author{Morten Hannibal Madsen\textsuperscript{2}}
\affiliation{\textsuperscript{1}Microsoft Quantum Materials Lab Copenhagen, 2800 Lyngby, Denmark}
\affiliation{\textsuperscript{2}Center for Quantum Devices, Niels Bohr Institute, University of Copenhagen, 2100 Copenhagen, Denmark \\ 
\textsuperscript{*}These authors contributed equally \\ 
\textsuperscript \textdagger Corresponding author: krogstrup@nbi.dk}

\author{Rawa Tanta\textsuperscript{2}}
\affiliation{\textsuperscript{1}Microsoft Quantum Materials Lab Copenhagen, 2800 Lyngby, Denmark}
\affiliation{\textsuperscript{2}Center for Quantum Devices, Niels Bohr Institute, University of Copenhagen, 2100 Copenhagen, Denmark \\ 
\textsuperscript{*}These authors contributed equally \\ 
\textsuperscript \textdagger Corresponding author: krogstrup@nbi.dk}

\author{Gunjan Nagda\textsuperscript{1,2}}
\affiliation{\textsuperscript{1}Microsoft Quantum Materials Lab Copenhagen, 2800 Lyngby, Denmark}
\affiliation{\textsuperscript{2}Center for Quantum Devices, Niels Bohr Institute, University of Copenhagen, 2100 Copenhagen, Denmark \\ 
\textsuperscript{*}These authors contributed equally \\ 
\textsuperscript \textdagger Corresponding author: krogstrup@nbi.dk}

\author{Sergej Schuwalow\textsuperscript{1,2}}
\affiliation{\textsuperscript{1}Microsoft Quantum Materials Lab Copenhagen, 2800 Lyngby, Denmark}
\affiliation{\textsuperscript{2}Center for Quantum Devices, Niels Bohr Institute, University of Copenhagen, 2100 Copenhagen, Denmark \\ 
\textsuperscript{*}These authors contributed equally \\ 
\textsuperscript \textdagger Corresponding author: krogstrup@nbi.dk}

\author{Peter Krogstrup\textsuperscript{1,2,\textdagger}}
\affiliation{\textsuperscript{1}Microsoft Quantum Materials Lab Copenhagen, 2800 Lyngby, Denmark}
\affiliation{\textsuperscript{2}Center for Quantum Devices, Niels Bohr Institute, University of Copenhagen, 2100 Copenhagen, Denmark \\ 
\textsuperscript{*}These authors contributed equally \\ 
\textsuperscript \textdagger Corresponding author: krogstrup@nbi.dk}

\date{\today}  
\email{Corresponding author: krogstrup@nbi.dk}

\begin{abstract}

Selective area growth (SAG) of semiconductors is a scalable method for fabricating gate-controlled quantum platforms. This letter reports on the adatom diffusion, incorporation, and desorption mechanisms that govern the growth rates of SAG nanowire (NW) arrays. We propose a model for the crystal growth rates that considers two parameter groups: the crystal growth control parameters and the design parameters. Using GaAs and InGaAs SAG NWs as platform we show how the design parameters such as NW pitch, width, and orientation have an impact on the growth rates. We demonstrate that by varying the control parameters (i.e. substrate temperature and beam fluxes) source, balance, and sink growth modes may exist in the SAG selectivity window. Using this model, we show that inhomogeneous growth rates can be compensated by tuning the design parameters.
\end{abstract}

\maketitle            

\section{Introduction}

One-dimensional semiconductor nanowires (NWs) have the potential to become the host platform of future quantum information technologies \cite{stanescu2013majorana, karzig2017scalable, plugge2017majorana, lutchyn2010majorana}. Among the different crystal growth techniques of semiconductor NWs \cite{wagner1964vapor, kato1994selective, johar2020universal, ambrosini2011self, khan2020highly}, Selective Area Growth (SAG) of in-plane III-V NWs using Molecular Beam Epitaxy (MBE) is a method for synthesizing scalable gate-controlled one-dimensional quantum electronics \cite{heiss2008growth, desplanque2014influence}. In particular, the design flexibility allows for arbitrary device architectures including networks of quantum dots and NWs. 

Recently, in-plane III-V NW arrays have gained attention and have been demonstrated with a variety of materials, shapes and dimensions \cite{krizek2018field, vaitiekenas2018selective, anselmetti2019end, desplanque2018plane, fahed2016selective, fahed2015impact, tutuncuoglu2015towards, lee2019selective, friedl2018template, aseev2018selectivity}. Optimising SAG crystal growth for the functionality and quality of such quantum structures implies optimizations of the morphology \cite{fahed2015impact}, composition \cite{bacchin1998dependence}, crystal disorder \cite{wimmer2011quantum, flensberg2010tunneling, yang2017revealing} and strain uniformity \cite{fahed2017threading}. This makes it necessary to control the incorporation rates with high precision in order to optimize the performance of quantum devices \cite{aseev2019ballistic, het2020plane}.

Here we present a study of adatom incorporation during crystal growth of in-plane GaAs and InGaAs SAG NWs, grown by MBE on GaAs(001) and InP(001) substrates. A silicon dioxide (SiO$_2$) mask is used to define NWs on the substrates \cite{krizek2018field}. We discuss two groups of parameters that affect SAG NW growth rates: \textit{growth control} parameters, namely substrate temperature ($T_{sub}$) and beam fluxes ($f_i$), and the \textit{NW design} such as the width ($w$), interwire pitch ($p$), and in plane crystallographic orientation $[hkl]$. These dependencies are of importance for the design of reproducible arrays of NWs.

\section{Adatom kinetics}

A crystal growth by MBE is facilitated by incoming beam fluxes of growth species that impinge and get adsorbed on the substrate surface \cite{sugaya1992selective, okamoto1987selective, okamoto1989substrate}. We describe this mechanism by transition state kinetics of the adatoms, where the transition rates $\Gamma_{\alpha\beta}$ ($\alpha$ and $\beta$ denote the initial and final state, respectively) are limited by effective kinetic barriers \cite{krogstrup2013}. To achieve SAG, the adatoms on the mask must either desorb or diffuse to the exposed crystal areas. The crystal growth rate is highly dependent on $T_{sub}$ and $f_i$, as well as the surface state parameters, e.g. activation energies for adatom desorption, surface diffusion, nucleation and incorporation \cite{aseev2018selectivity}. Following the continuum kinetics approach in ref.\cite{krogstrup2013} and ignoring adatom chemical potential variations, the transition rates can be described by the Arrhenius equation $\Gamma_{\alpha\beta} \propto \rho_{\alpha} \exp\left(-\frac{\delta g_{\alpha\beta}}{k_B T_{sub}}\right)$, where $k_B$ is the Boltzmann constant, $\rho_{\alpha}$ is the adatom density in the initial state, and $\delta g_{\alpha\beta}$ is the effective activation energy for the transition.

\begin{figure}[b]
    \centering
    \includegraphics[width=0.48\textwidth]{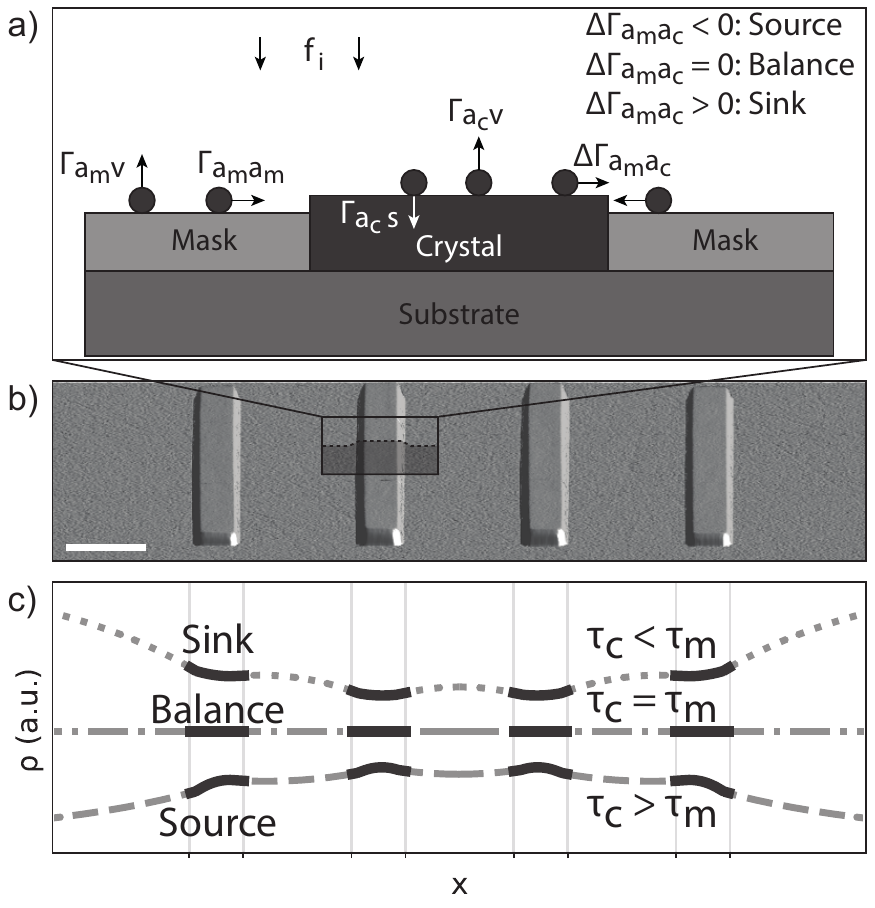}
    \vspace{-10 pt}
    \caption{(a) Schematic of adatom transitions during SAG. (b) AFM image of four parallel GaAs NWs oriented along the $[1\bar{1}0]$ crystal orientation on a GaAs (001) substrate. Scale bar is 500 nm. (c) Three types of solutions to the simulation of coupled diffusion equations describing the adatom density on four parallel NWs assuming infinite length. The dashed and solid lines are the adatom density on mask and crystal regions, respectively.}
    \label{fig:fig1}
\end{figure}

Figure \ref{fig:fig1}a sketches the different types of adatom transitions which take place during SAG: adatom diffusion on either the mask surface ($\Gamma_{a_ma_m}$), the growing crystal surface ($\Gamma_{a_ca_c}$) or across a mask-crystal boundary ($\Gamma_{a_ma_c}$); adatom incorporation into the crystal solid phase ($\Gamma_{a_cs}$), or via nucleation to solid phase on the mask ($\Gamma_{a_ms}$); or adatom desorption from the mask ($\Gamma_{a_mv}$) and crystal ($\Gamma_{a_cv}$) to vapour. All transition rates in this study are effective rates describing the mean properties of the transitions, e.g.  $\Gamma_{a_cs}$ describes both the nucleation limited transitions as well as potential single atomic barriers for incorporation. Due to the geometry, the mask-crystal boundary is 1D-like (linear) and the boundary of the $\Gamma_{a_cs}$, $\Gamma_{a_cv}$, $\Gamma_{a_ms}$ and $\Gamma_{a_mv}$ transitions is 2D-like (surface).

The total current of adatoms of species \textit{i} being incorporated in a NW segment of length $l$ and width $w$ ($l \gg w$) is given by adatom conservation: 
\begin{equation}
    I_{a_cs,i} = (f_i - \Gamma_{a_cv,i})\cdot w\cdot l + (\Gamma_{a_ma_c,i} - \Gamma_{a_ca_m,i})\cdot 2l,
    \label{eq:NW_inc_current}
\end{equation}
assuming no substrate decomposition $\Gamma_{sa_c}=0$. From mass conservation, the incorporation rate on the SAG NW can therefore be written as
\begin{equation}
     \Gamma_{a_cs} = \sum_{i} f_i - \Gamma_{a_cv,i} + 2\frac{\Gamma_{a_ma_c,i} - \Gamma_{a_ca_m,i}}{w} .
     \label{eq:NW_inc_rate}
\end{equation}
Then, the crystal volume growth rate is $\Gamma_{a_cs}\cdot\Omega$  where $\Omega$ is the volume of a III-V atomic pair. While $f_i$ is a controlled parameter, the desorption term $\Gamma_{a_cv,i}$ is highly dependent on $T_{sub}$. Thus, if the desorption from the crystal can be ignored for a given $T_{sub}$, i.e. $\Gamma_{a_cv}=0$, the relevant term for controlling the growth rate is the flux across the mask-crystal boundary,  
\begin{equation}
    \Delta\Gamma_{a_ma_c,i} = \Gamma_{a_ma_c,i} - \Gamma_{a_ca_m,i},
\end{equation}
where the forward flux $\Gamma_{a_ma_c,i}$ is the flux of adatoms to the crystal collected from the mask and the backward flux $\Gamma_{a_ca_m,i}$ is the flux of adatoms to the mask collected from the crystal.
We define the growth mode as \textit{source} if $\Delta\Gamma_{a_ma_c,i}<0$, \textit{sink} if $\Delta\Gamma_{a_ma_c,i}>0$, and \textit{balance} if $\Delta\Gamma_{a_ma_c,i}=0$. 

To simulate the adatom fluxes in this system, we start by simplifying the adatom diffusion problem to one dimension by considering only the transversal direction \textit{x} of an array of parallel NWs of infinite $l$. The steady state adatom diffusion equation for each surface \textit{j} can be written as  
\begin{equation}
    D_j \frac{\partial^2 \rho_j(x)}{\partial x^2} + f - \Gamma_{a_jv}(x) - \Gamma_{a_js}(x) = {0} ,
    \label{eq:steady_state}
\end{equation}
where the mask and crystal surfaces $j$ are coupled via boundary conditions for the particular design in question. We consider four NWs in parallel with symmetry at the midpoint between the two inner NWs ($x=0$) and at $x=\infty$, i.e. $\frac{\partial \rho_{a_j}}{\partial x}=0$ (see Supplementary S1). Figure \ref{fig:fig1}b is an Atomic Force Microscopy (AFM) image of a typical array of four NWs used in this work.
At the boundaries, we assume continuity, $\rho_{a_c}=\rho_{a_m}$, and mass conservation, $D_c\frac{\partial \rho_{a_c}}{\partial x}=D_m\frac{\partial \rho_{a_m}}{\partial x}$. 
Since the effective incorporation rate is proportional to the adatom density, $\Gamma_{a_cs} \propto \rho_{a_c} \exp\left(-\frac{\delta g_{a_cs}}{k_B T}\right)$, we are interested in solving $\rho_{a_c}$ as a measure of the SAG growth rate. 
The model exhibits three general types of solutions, shown in Figure \ref{fig:fig1}c: 
sink, balance and source growth modes. If there is no adatom desorption from the crystal $\Gamma_{a_cv}=0$, the sink (source) mode implies that the NW growth rate is higher (lower) than the calibrated corresponding planar growth rate (2D-like growth with no mask), due to an inhomogeneous flux of adatoms $\Gamma_{a_ma_c}$ at the boundary. The balance mode implies that the NW growth rate is equal to the corresponding planar growth rate, and importantly, it is \textit{independent of NW design}. 
These distinct growth modes have implications for the design of NW patterns. In the following, we will explore SAG growth rates of GaAs and InGaAs based NWs grown on GaAs and InP substrates with a SiO$_2$ mask and illustrate how growth modes and growth rates can be identified and quantified.

\section{Experimental methods}

The structures under study consist of individual NWs and arrays of NWs, with varying pitch and width. The NWs are 14 \textmu m long and to avoid influence from the ends, the measure of incorporation is only considered in the central region of the NW (see Supplementary S2). The employed SiO$_2$ mask fabrication flow and crystal growth concept by MBE are described in references \cite{krizek2018grown} and \cite{krizek2018field}, respectively. Supplementary S5 contains information about the typical roughness of mask and substrate prior MBE growth. The MBE beam fluxes are calibrated to the corresponding planar growth rates using Reflection High Energy Electron Diffraction (RHEED) oscillations under conditions where desorption of group III can be ignored \cite{aseev2018selectivity}. We calibrate the V:III 1:1 flux ratio with the surface reconstruction change procedure, using RHEED on GaAs(100) substrates \cite{neave1983dynamics, daweritz1990reconstruction}. Temperature is measured with pyrometer, which is calibrated with GaAs oxide desorption \cite{guillen2000gaas}. After growth, the SAG NW volume in single layer growths is measured by AFM and the cross sectional area on the multi-layer sample is measured by cross sectional Transmission Electron Microscopy (TEM). We define the NW growth rate, $\Gamma_{inc}$, as the measured crystal volume divided by the volume of a NW section with the same $w$ and $l$ from the equivalent planar growth used for the flux calibration. $\Gamma_{inc}$ is a measure of the amount of material incorporated in a NW compared to the 2D growth and hence the effect on material incorporation caused by the mask-crystal interface term $\Delta\Gamma_{a_ma_c}$ from eq. \ref{eq:NW_inc_rate}. The mask selectivity measurements are performed in areas with no mask openings, to avoid any influence from the $\Delta\Gamma_{a_ma_c}$ term.

\section{Results and discussions}

\begin{figure*}
    \centering
    \includegraphics[width=1\linewidth]{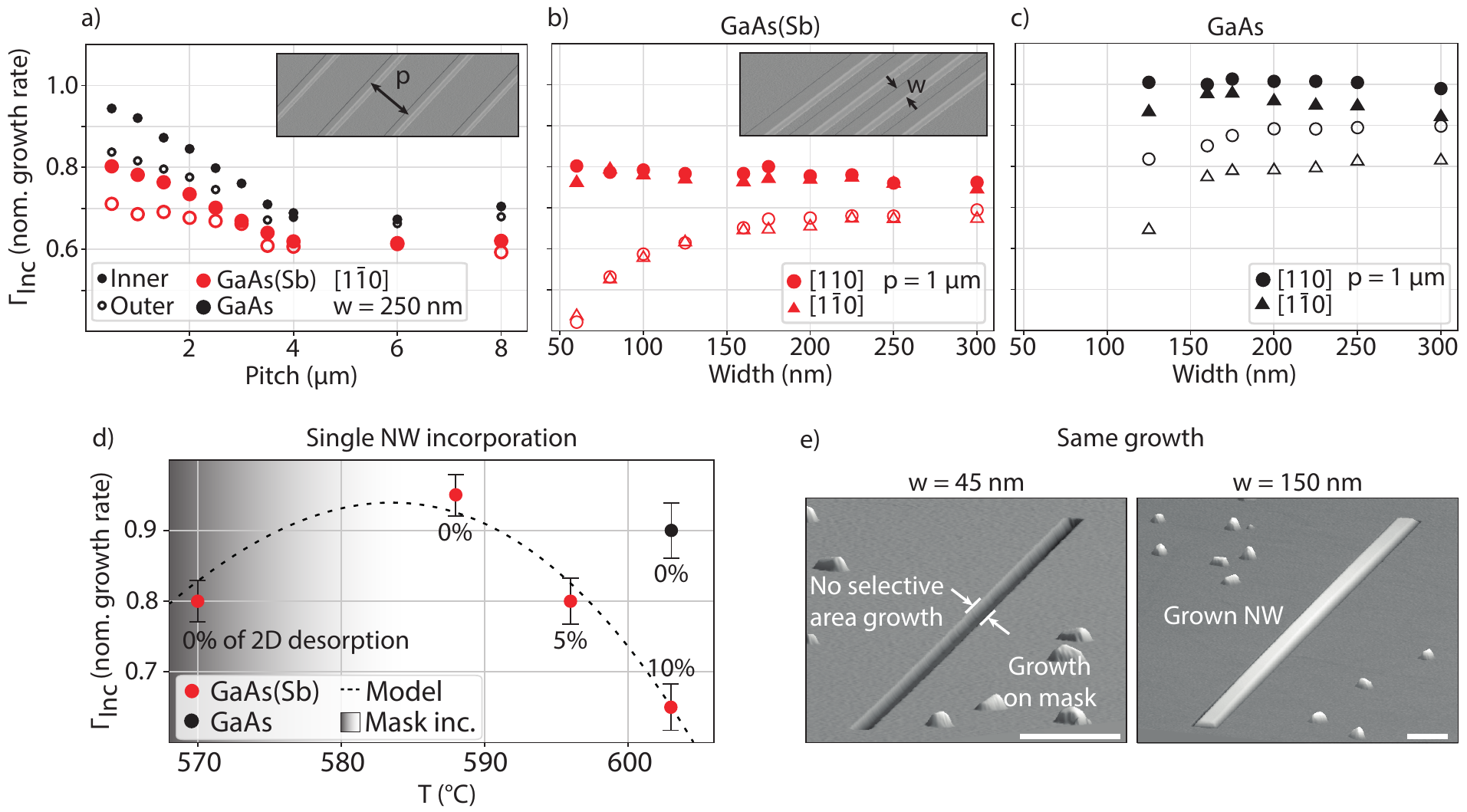}
    \caption{NW Incorporation rate dependence on design parameters, pitch, width, and $[hkl]$. Incorporation rates in a)-c) are measured in units of nominal growth rate of GaAs and GaAs(Sb). All three plots share the same y-scale. Filled (open) points indicate incorporation rates measured on the inner (outer) NWs. GaAs(Sb) is shown by red symbols, and GaAs by black symbols. a) Incorporation rates of inner and outer NWs in a 4 NWs array, as a function of NW $p$. The inset shows an AFM image of an example array. b) and c) are the incorporation rates of GaAs(Sb) and GaAs, respectively, as a function of $w$ and for $[110]$ and $[1\bar{1}0]$ oriented NWs. The inset in b) is an AFM image of an example array. d) Incorporation of isolated NWs as a function of $T_{sub}$ indicated by black (red) symbols, for GaAs (GaAs(Sb)). The dashed line is extracted from the model based on eq. \ref{eq:NW_inc_rate}, highlighting a maximum incorporation near 583 °C. The number below each data point corresponds to the percentage of desorbed material from the crystal $\Gamma_{a_cv}$, measured on large mask openings. The blurred grey background represent the transition from $\Gamma_{a_ms}=0$ (white) to $>0$ (grey). e) Example of individual isolated 45 and 150 nm wide NWs, both from the same GaAs(Sb) growth. The source effect overrides the growth inside the 45 nm trench due to its width limitation, whereas the 150 nm grows as expected. Scale bars are 200 nm.}
    \label{fig:fig2}
\end{figure*}

We start by examining the growth rates in arrays of four parallel NWs, shown in Figures \ref{fig:fig2}a and b insets, as a function of design parameters $p$, $w$ and $[hkl]$. 
We determine the growth mode by comparing incorporation rates between the two inner and two outer NWs. Figure \ref{fig:fig2}a-c shows the mean incorporation rate of pure GaAs and Sb surfactant-aided GaAs(Sb) NWs grown on a GaAs(001) substrate at $T_{sub}$ of 603 °C. The Ga flux corresponds to a planar GaAs growth rate of 0.1 monolayers (ML)/s, under As rich conditions (see Supplementary S4 for recipe details). The reason for the selection of these two materials is their use as buffer layers before the growth on InAs transport channels, due to its beneficial effect of crystal defect reduction at the InAs interface \cite{krizek2018field}.
In Figure \ref{fig:fig2}a the mean incorporation rates of inner and outer NWs is plotted as a function of $p$ for $[1\bar{1}0]$ orientated NWs of $w=250$ nm. The data reveals a decrease in incorporation rates with increasing $p$ until it saturates at around 4~\textmu m for both GaAs and GaAs(Sb) to 0.7 and 0.6 of the nominal incorporation rate, respectively. We note that the $[1\bar{1}0]$ NWs exhibit different faceting with and without Sb surfactant, with $(001)$ vertical and $\{113\}$ side predominant facets, respectively \cite{beznasyuk2021role} (more details about faceting in Supplementary S2). The different faceting can affect the total incorporation of the NW due to Ga adatom diffusion length $\lambda_{Ga,c}$ anisotropy on GaAs(001) \cite{yamamura2005influence}. However, in the $p$ study we consider this effect negligible since $w \ll \lambda_{Ga,c_{[110]}}$ and $\lambda_{Ga,c_{[1\bar{1}0]}}$ \cite{ roehl2010binding, minkowski2015diffusion}. The facet time evolution of  GaAs NWs from initial (001) to \{113\} is not considered because the initial stages of the growth are dominated by the diffusion on the original (001) substrate and the NWs reach the \{113\} fully grown facets at the end of the growth process. The incorporation rate of pure GaAs arrays approaches the nominal value (i.e. $\Gamma_{inc}=1$) at small $p$, as expected if the desorption from the crystal is negligible, i.e. $\Gamma_{a_cv,Ga}\approx0$. By contrast, GaAs(Sb) NWs have a have lower incorporation rates but with the same overall trend.
This general downwards shift of $\sim$ $10~\%$ in the incorporation curve for the Sb surfactant compared with the pure case can be explained by two reasons: a result of non-reactive surfactants nature, which decrease the number of incorporation sites for adatoms \cite{tournie1995surfactant,oh1996kinetic} and the higher facet roughness of \{113\} compared to (001).
The decrease in incorporation with pitch implies that the growth is in the source mode ($\Delta\Gamma_{a_ma_c} < 0$). As a consequence, an increasing pitch implies a decreasing number of adatoms being shared between neighbouring NWs before they are desorbed from the mask. In the regime of significantly large pitch ($p \gg \lambda_{Ga,m}$)\cite{heiss2008growth, okamoto1993selective}, the sourcing of adatoms between NWs can be ignored. All NWs in the array grow at the same rate and can be considered decoupled from each other. In this regime, the amount of material incorporated by a NW compared to the nominal growth rate is a direct measure of the source mode strength for the given growth conditions.

\begin{figure}
    \centering
    \includegraphics[width=\linewidth]{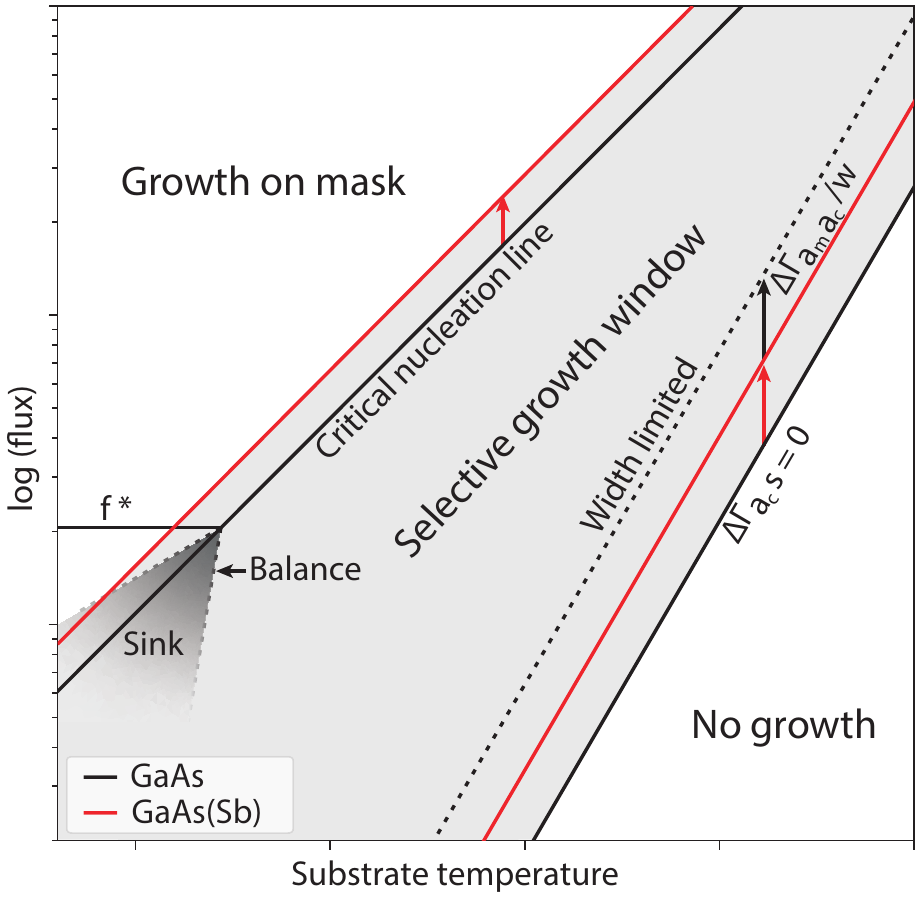}
    \caption{Sketch of SAG growth window and the implications of the source effect. The transition from solid black to solid red lines indicated by the red arrows is caused by the use of Sb surfactant during growth, shifting the $\Gamma_{a_cs}=0$ and $\Gamma_{a_ms}=0$ lines towards lower T. The transition from solid red to dotted black on the $\Gamma_{a_cs}=0$ line is caused by the width limitation of the design, additionally shifting it towards lower T. A region of sink growth mode may exist at low f and T, delimited by a line that would be balanced growth mode. Above the sink region maximum flux f$^*$, all growths are expected to be in source mode.}
    \label{fig:fig3}
\end{figure}

While the $p$ dependence is used to study the desorption limited $\lambda_{Ga,m}$, the $w$ dependence can be used to study the incorporation limited $\lambda_{Ga,c}$. Figure \ref{fig:fig2}b and c show the incorporation rate as a function of $w$, with and without Sb surfactant respectively, for both [110] and $[1\bar{1}0]$ oriented NW arrays with $p=1$ \textmu m. 
As shown in Figure \ref{fig:fig2}b, the GaAs(Sb) growth rate is independent of $[hkl]$ and even independent of $w$ for the inner NWs. However, for the outer NWs there is a decrease in the growth rate with decreasing $w$. The outer NW growth rate dependence on $w$ is consistent with equation \eqref{eq:NW_inc_rate}; as $w$ increases the sourcing effect from $\Delta\Gamma_{a_ma_c}$ becomes negligible for the given NW, and the incorporation rate converges towards $\Gamma_{a_cs} = f_i - \Gamma_{a_cv,i}$. If $\Gamma_{a_cv,Ga}\approx0$ and the width approaches $w \gg \lambda_{Ga,c}$ the mean incorporation rate will converge towards the nominal growth rate for all NWs in the array, ignoring the effect of the surfactant. For increasing widths the growth rate converges towards 0.7 - 0.8 of the nominal growth rate, which can be explained with a longer $\tau_{Ga,c}$ due to the role of the surfactant and therefore a higher $\Gamma_{a_ca_m}$. On the other hand, the apparent $w$ independence on incorporation for the inner GaAs(Sb) NWs is not obvious. We speculate that this apparent independence of $\Gamma_{inc}$ is due to a compensation on growth rates. The inner NWs get more sourced adatoms from its neighbours,  as the outer NWs incorporate less at smaller $w$.
The $\Gamma_{inc}$ difference between both inner and outer NWs on both directions with $w = 150 - 300$ nm slowly decreases and it is expected to merge at larger $w$.
In Figure \ref{fig:fig2}c, there is a clear dependence on $[hkl]$ and the incorporation rate of GaAs is more efficient on the crystal with $\Gamma_{inc}$ closer to 1. 
As the NWs grow, the faceting evolves differently depending on the NW orientation, which means that $\lambda_{Ga,c}$ also changes during growth. Thus, the crystal surface parameters can be dynamic in nature. The NWs oriented along $[1\bar{1}0]$ form dominating $\{113\}$ facets, while the $[110]$ oriented NWs preserve the $(001)$ top facet. As such, the $[1\bar{1}0]$-oriented NWs exhibit a stronger source effect because the longer lifetime results in a lower incorporation rate, and therefore a more negative $\Delta\Gamma_{a_ma_c}$ compared to the NWs oriented along $[110]$. This is consistent with the findings in reference \cite{sato2004growth} which show the incorporation rate of Ga on $\{113\}$ GaAs facets is slower than on $(001)$. As we are measuring only mean growth rates, for simplicity we also assume constant surface state parameters for the modelling, and any change in faceting during the growth is not considered.

The growths discussed in the previous paragraphs exhibit a source behavior. To answer if it is possible to manipulate the strength of the source effect $\Delta\Gamma_{a_ma_c}$, and potentially achieve balanced and sink growth modes, we grow four identical GaAs(Sb) samples where only $T_{sub}$ is varied between 570 °C and 603 °C at a nominal growth rate of 0.1~ML/s with a V/III ratio of 9. As shown in reference \cite{yokoyama1989low}, the reduction of $T_{sub}$ leads to an exponential reduction of $\Gamma_{a_mv}$, decreasing $\Gamma_{a_ca_m}$ since the adatom density will be higher on the mask. Incorporation rates are measured on isolated NWs of $w = 250$ nm for each growth, and plotted in Figure~\ref{fig:fig2}d as a function of $T_{sub}$. GaAs(Sb) NWs initially increase the incorporation when temperature is reduced, compared to the growths at 603 °C discussed in Figures~\ref{fig:fig2}a-c. The number near each point in Figure~\ref{fig:fig2}d is the 2D desorption ($\Gamma_{a_cv}$) in percents of nominal growth rate . The dashed line is the model prediction of incorporation for the GaAs(Sb) samples, based on the adatom conservation model from equation \ref{eq:NW_inc_rate} (see Supplementary S1) and highlighting a maximum in incorporation around 583 °C for the given growth rate of 0.1 ML/s. None of the samples measured in this series reach nominal incorporation. There are two independent reasons. First, for the two highest temperature samples, the crystal desorption, $\Gamma_{a_cv}$, on large mask openings (i.e. 2D-like) is non-negligible and 10\% and 5\% at 603 °C and 595 °C, respectively (see Supplementary S8 for 2D desorption measurements). Second, the source effect term $\Delta\Gamma_{a_ma_c}$ reduces the incorporation further at lower temperatures as it gets more negative, as seen in the sample grown at 570 °C. This is understood by the decrease of adatom density on the mask due to nucleation of parasitic crystals $\Gamma_{a_ms}$ near the NWs, leading to an increase of the transition $\Gamma_{a_ca_m}$. The parasitic growth on the mask is marked in Figure~\ref{fig:fig2}d with the blurred grey background, with a transition happening between 570 °C and 588 °C. Figure~\ref{fig:fig2}e shows two AFM images of individual GaAs(Sb) NWs from the same growth, with different $w$, and separated 50 \textmu m from other mask opening. For the 45 nm wide NW, due to the small $w$, the term $\Delta\Gamma_{a_ma_c}/w$ from eq. \ref{eq:NW_inc_rate} is negative enough to override the growth on the substrate due to the dominating $\Gamma_{a_ca_m}$, whereas parasitic growth can nucleate on the mask. This would be an extreme case where the design parameter $w$ induces a negative growth rate for the NW at growth conditions that otherwise induce growth in wider mask openings, as shown in the 150 nm wide NW.


\begin{figure*}
    \centering
    \includegraphics[width=1\linewidth]{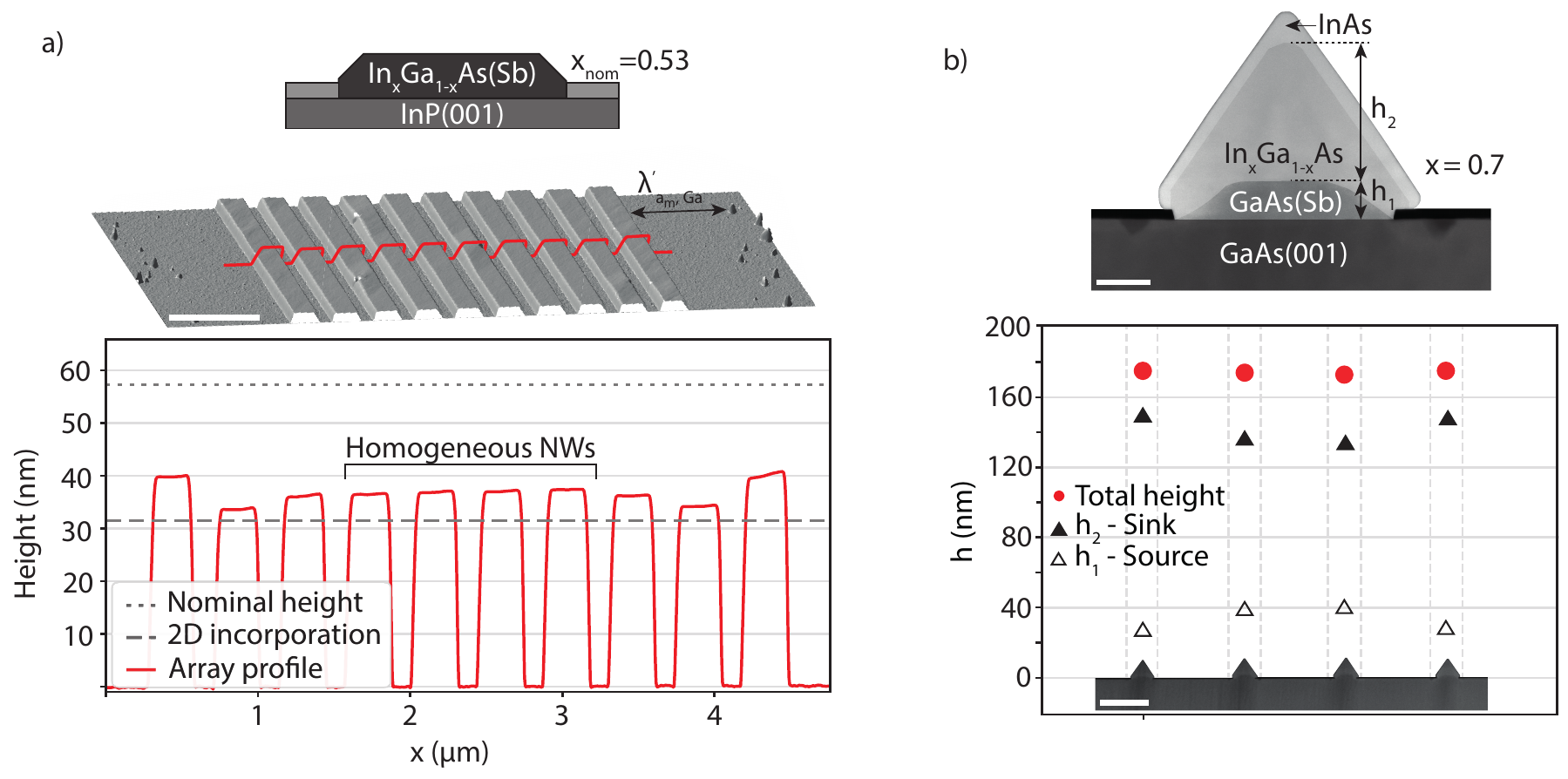}
    \caption{Engineering arrays of NWs with homogeneous height. a) Sketch and AFM image of an array of In$\textsubscript{0.53}$Ga$\textsubscript{0.47}$As NWs grown on InP(001). In the bottom, AFM line scan across the NWs showing simultaneously both source and sink growth modes, generated by the different group III adatoms. The central NWs of the array form a set of NWs with homogeneous height, due to the formation of an adatom density saturation region. b) ADF-STEM cross sectional image of multi-layer buffer NWs (top) and low magnification of the NW array (bottom). Material contrast is highlighted for visualization purposes. The plot shows the height of the 2 buffers for each of the 4 NWs in the array, summing up to the same height due to the balanced effect of source and sink growth modes. Scale bars are 1 \textmu m in a), 50 nm in b) (top) and 500 nm in b) (bottom).}
    \label{fig:fig5}
\end{figure*}

The selectivity window for SAG NWs describes suitable growth conditions in the temperature-flux space, as studied by Aseev et al. in ref. \cite{aseev2018selectivity} for GaAs (001) substrates and SiO$_2$ mask. Based on our results, we schematically introduce in Figure~\ref{fig:fig3} the effect on the lower (desorption from the crystal, $\Gamma_{a_cv}$) and upper (nucleation on the mask, $\Gamma_{a_ms}$) boundaries caused by the Sb surfactant, and the NW design limitation on $w$. The solid black curves in Figure~\ref{fig:fig3} represent the upper and lower boundaries and they confine the region in the $f$ - T space where the growth is selective. From the results presented in Figure \ref{fig:fig2}a-d, the effect on these boundaries due to the addition of Sb surfactant is the shift to lower temperatures (or higher fluxes), sketched with solid red lines. It is unclear if the shift is identical for both boundaries since the transition rates defining them might be affected differently by Sb. The growth limitation on the design parameter $w$ shown in Figure \ref{fig:fig2}e has an implication on the lower boundary by additionally shifting the curve towards lower temperatures by a factor proportional to the strength of the source effect, $\Delta\Gamma_{a_ma_c}/w$. This shift can cause the lower boundary to cross the upper boundary, effectively closing the SAG window as shown in Figure \ref{fig:fig2}e. 
We emphasize that for previously reported SAG $f$ - T conditions where the $w$ limitation in source mode was not taken into account, it is possible to override the growth.
Next, as demonstrated in Figure \ref{fig:fig2}d, the source effect can be reduced by decreasing the growth temperature, until the point where nucleation on the mask starts. Reducing $f$ while still growing selectively would allow to neutralize the source effect by shifting the incorporation curve in Figure \ref{fig:fig2}d upwards until $\Gamma_{inc}=1$. In Figure \ref{fig:fig3} we speculate the appearance of a sink region at low $f$ and T where $\Delta\Gamma_{a_ma_c}>0$ and whose boundary gives balance growth mode independent of NW design. The sink window is expected to extend symmetrically beyond the upper boundary towards lower temperatures until the adatom density is sufficiently reduced by nucleation on the mask, entering the source growth mode once more. This localized region in the $f$ - T selectivity map would imply the existence of a critical flux $f^*$ above which it would not be possible to achieve a balanced growth. Further exploration is needed at lower group III fluxes and temperatures of the SAG window in order to demonstrate a balanced growth mode and the existence of the sink effect region for binary materials.


\section{Engineering of III-V ternary materials}

Fabrication of SAG NWs for quantum electronic devices usually requires the growth of multi-stack buffer layers to minimize the generation of defects that degrade electronic properties \cite{krizek2018field}. Specifically, the growth of In$_x$Ga$_{1-x}$As buffer layers between the substrate and the InAs transport channel has been demonstrated to be beneficial for the strain relaxation of InAs \cite{beznasyuk2021role}. Using this approach, in Figure \ref{fig:fig5} we show two independent methods of engineering arrays of NWs with constant height based on the presented source and sink growth regimes. 

Figure \ref{fig:fig5}a shows an array of lattice matched In$_{0.53}$Ga$_{0.47}$As NWs grown on InP(001). The three outermost NWs are under the influence of the nearby mask, whereas the four middle NWs have a constant height. This approach to grow several NWs homogeneously can be extended by increasing the number of NWs in the array, generating a central region in each array where adatom density is constant. In Figure \ref{fig:fig5}a, the outermost NWs are the highest in the array. This is caused by the different behavior of group III adatoms of the In$_{0.53}$Ga$_{0.47}$As ternary alloy, as opposed to the pure source case from Figure \ref{fig:fig2}a. For the growth $T_{sub}=508$ °C and for the given fluxes, we speculate that Ga adatoms are being incorporated in the sink regime of their SAG window whereas the In adatoms are in source regime simultaneously (see Supplementary S2 for growth conditions). This explains a local sink behavior for the outer NWs and a general source effect for the inner NWs compared to the second and third outermost ones.

Another approach of growing reproducible structures is presented in Figure \ref{fig:fig5}b. The NWs consist of two buffer layers of GaAs(Sb) and In$_{0.7}$Ga$_{0.3}$As with an InAs transport channel, grown on a GaAs(001) substrate. The Annular Dark Field Scanning TEM on Figure \ref{fig:fig5}b shows the cross-sectional geometry and contrast between the different layers. Material contrast is highlighted for visualization purposes. The In$_{0.7}$Ga$_{0.3}$As composition is extracted by Electron Energy Loss Spectroscopy and its growth temperature dependence and array position variation has been studied in depth in reference \cite{beznasyuk2021role}. The inset in the plot is the cross sectional lamella of the array in study via Focused Ion Beam milling. The white (black) triangles show the height of the first (second) buffers. Here, a combination of source growth mode from the first buffer and sink growth mode from the second buffer generates an array of NWs with constant height, where the subsequent InAs layer was grown on.

\section{Conclusion}
In summary, we have measured and analyzed the SAG growth rates of GaAs, GaAs(Sb) and In$_x$Ga$_{1-x}$As NWs on GaAs/SiO$_2$ patterned substrates. We show how the growth rates are dominated by the effective flux of adatoms across the mask to crystal areas, $\Delta\Gamma_{a_ma_c}$, where the sign of $\Delta\Gamma_{a_ma_c}$ determines whether the growth is in source (negative), balanced (neutral) or sink (positive) growth mode. The growth mode is determined by measuring the growth rate dependence on the variables: NW array pitch, position, width, crystallographic orientation, and chemical composition. With the growth conditions used in this study, GaAs and GaAs(Sb) grow consistently in source mode while In$_x$Ga$_{1-x}$As grows effectively in sink mode. We demonstrate the possibility of growing reproducible NWs by two different approaches: tuning the growth mode of each group III species on buffer layer stacks and by increasing the number of NWs in the array, creating uniform incorporation conditions for adjacent NWs.

\section{Acknowledgement}

The research was supported by Microsoft Quantum Initiative and the European Research Council under the Horizon 2020 research and innovation program, starting grant HEMs-DAM, agreement No. 716655. The authors thank Mohana Rapjalke for valuable discussions and technical support. M.E.C. thanks Elisabetta Fiordaliso for support in microscopy.

\bibliography{bibliography}

\end{document}


\title{Selective Area Growth Rates of III-V Nanowires}

\author{Martín Espiñeira Cachaza\textsuperscript{1,2,*}}
\affiliation{\textsuperscript{1}Microsoft Quantum Materials Lab Copenhagen, 2800 Lyngby, Denmark}
\affiliation{\textsuperscript{2}Center for Quantum Devices, Niels Bohr Institute, University of Copenhagen, 2100 Copenhagen, Denmark \\ 
\textsuperscript{*}These authors contributed equally \\ 
\textsuperscript \textdagger Corresponding author: krogstrup@nbi.dk}

\author{Anna Wulff Christensen\textsuperscript{1,2,*}}
\affiliation{\textsuperscript{1}Microsoft Quantum Materials Lab Copenhagen, 2800 Lyngby, Denmark}
\affiliation{\textsuperscript{2}Center for Quantum Devices, Niels Bohr Institute, University of Copenhagen, 2100 Copenhagen, Denmark \\ 
\textsuperscript{*}These authors contributed equally \\ 
\textsuperscript \textdagger Corresponding author: krogstrup@nbi.dk}

\author{Daria Beznasyuk\textsuperscript{1,2}}
\affiliation{\textsuperscript{1}Microsoft Quantum Materials Lab Copenhagen, 2800 Lyngby, Denmark}
\affiliation{\textsuperscript{2}Center for Quantum Devices, Niels Bohr Institute, University of Copenhagen, 2100 Copenhagen, Denmark \\ 
\textsuperscript{*}These authors contributed equally \\ 
\textsuperscript \textdagger Corresponding author: krogstrup@nbi.dk}

\author{Tobias S\ae rkj\ae r\textsuperscript{1,2}}
\affiliation{\textsuperscript{1}Microsoft Quantum Materials Lab Copenhagen, 2800 Lyngby, Denmark}
\affiliation{\textsuperscript{2}Center for Quantum Devices, Niels Bohr Institute, University of Copenhagen, 2100 Copenhagen, Denmark \\ 
\textsuperscript{*}These authors contributed equally \\ 
\textsuperscript \textdagger Corresponding author: krogstrup@nbi.dk}

\author{Morten Hannibal Madsen\textsuperscript{2}}
\affiliation{\textsuperscript{1}Microsoft Quantum Materials Lab Copenhagen, 2800 Lyngby, Denmark}
\affiliation{\textsuperscript{2}Center for Quantum Devices, Niels Bohr Institute, University of Copenhagen, 2100 Copenhagen, Denmark \\ 
\textsuperscript{*}These authors contributed equally \\ 
\textsuperscript \textdagger Corresponding author: krogstrup@nbi.dk}

\author{Rawa Tanta\textsuperscript{2}}
\affiliation{\textsuperscript{1}Microsoft Quantum Materials Lab Copenhagen, 2800 Lyngby, Denmark}
\affiliation{\textsuperscript{2}Center for Quantum Devices, Niels Bohr Institute, University of Copenhagen, 2100 Copenhagen, Denmark \\ 
\textsuperscript{*}These authors contributed equally \\ 
\textsuperscript \textdagger Corresponding author: krogstrup@nbi.dk}

\author{Gunjan Nagda\textsuperscript{1,2}}
\affiliation{\textsuperscript{1}Microsoft Quantum Materials Lab Copenhagen, 2800 Lyngby, Denmark}
\affiliation{\textsuperscript{2}Center for Quantum Devices, Niels Bohr Institute, University of Copenhagen, 2100 Copenhagen, Denmark \\ 
\textsuperscript{*}These authors contributed equally \\ 
\textsuperscript \textdagger Corresponding author: krogstrup@nbi.dk}

\author{Sergej Schuwalow\textsuperscript{1,2}}
\affiliation{\textsuperscript{1}Microsoft Quantum Materials Lab Copenhagen, 2800 Lyngby, Denmark}
\affiliation{\textsuperscript{2}Center for Quantum Devices, Niels Bohr Institute, University of Copenhagen, 2100 Copenhagen, Denmark \\ 
\textsuperscript{*}These authors contributed equally \\ 
\textsuperscript \textdagger Corresponding author: krogstrup@nbi.dk}

\author{Peter Krogstrup\textsuperscript{1,2,\textdagger}}
\affiliation{\textsuperscript{1}Microsoft Quantum Materials Lab Copenhagen, 2800 Lyngby, Denmark}
\affiliation{\textsuperscript{2}Center for Quantum Devices, Niels Bohr Institute, University of Copenhagen, 2100 Copenhagen, Denmark \\ 
\textsuperscript{*}These authors contributed equally \\ 
\textsuperscript \textdagger Corresponding author: krogstrup@nbi.dk}

\date{\today}  
\email{Corresponding author: krogstrup@nbi.dk}

\maketitle

\onecolumngrid 

\section{S1. Adatom kinetics models} \label{S1}

\begin{figure}[h]
    \centering
    \includegraphics[width=0.58\textwidth]{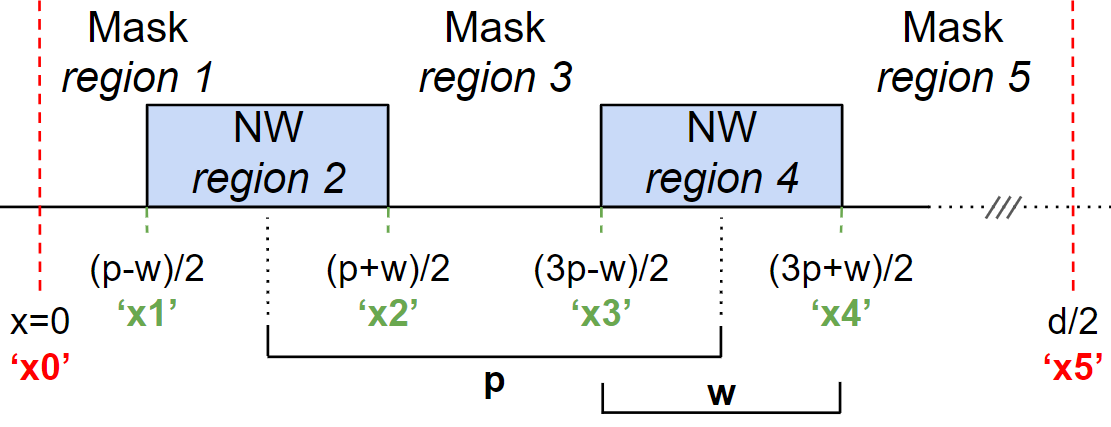}
    \caption{Sketch of the system geometry used in the coupled diffusion equations model for an array of 4 NWs. Points $x0$ and $x5$ are defined as symmetry points corresponding with the center of array and a point infinitely far from the NWs, respectively. At the boundaries $x1$, $x2$, $x3$, and $x4$ the adatom density is forced to fulfil mass conservation and continuity.}
    \label{fig:sup.fig1}
\end{figure}

\begin{figure}
    \centering
    \includegraphics[width=0.58\textwidth]{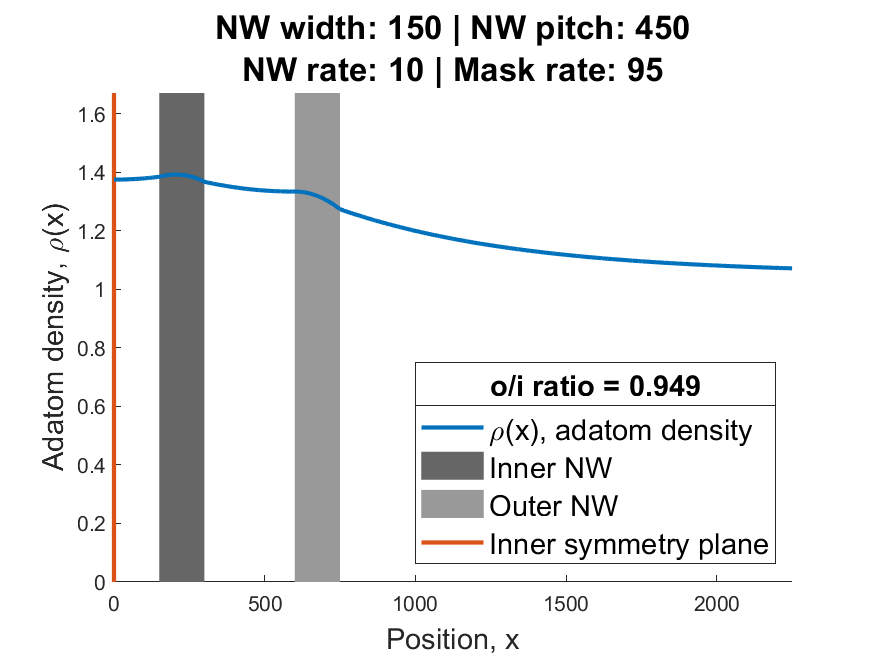}
    \caption{Adatom density simulation using the coupled diffusion equation model, for an array of 4 NWs in source growth mode.}
    \label{fig:sup.fig2}
\end{figure}

The adatom coupled diffusion equations model presented in equation 4 was computed in Matlab using the boundary value problem solver $bvp5c$. Figure \ref{fig:sup.fig1} shows an sketch of the geometry used in the model and the NW-mask boundaries where the adatom conservation equations are applied. A result of the model is presented in Figure \ref{fig:sup.fig2}, for the source case ($\tau_c$ > $\tau_m$) and the given array dimensions. For simplicity, the $\Delta\Gamma_{a_ma_c}$ term from the growth equation eq. 2 is expressed in terms of $\tau$.  The three solutions shown in Figure 1c are obtained by varying only the $\tau$ on the crystal while keeping the $\tau$ on the mask constant ($D$ is assumed the same across the surface for simplicity). It is considered no desorption from the substrate and no incorporation on the mask.

The adatom model presented in Figure 2d was computed analytically in Python. The growth equation used was:

\begin{equation}
    \begin{aligned}
     \Delta\Gamma_{a_ma_c} \propto \lambda_m - \lambda_c = \sqrt{D_{a_m}\tau_{a_m}} - \sqrt{D_{a_c}\tau_{a_c}} = \\
     =\sqrt{\frac{c_{D_m} \exp\left({\frac{-\delta g_{a_m a_m}}{k_B\ T}}\right)}{c_{a_m s} \exp \left({\frac{-\delta g_{a_m s}}{k_B\ T}}\right) +\ c_{a_m v}\exp \left( {\frac{-\delta g_{a_m v}}{k_B\ T}}\right)}} - \sqrt{\frac{c_{D_c} \exp\left({\frac{-\delta g_{a_c a_c}}{k_B\ T}}\right)}{c_{a_c s} \exp \left({\frac{-\delta g_{a_c s}}{k_B\ T}}\right) +\ c_{a_c v}\exp \left( {\frac{-\delta g_{a_c v}}{k_B\ T}}\right)}}
\end{aligned}
\label{eq:model_Fig2d}
\end{equation}

where the $\delta g$ are the activation energies for the respective adatom state transition and $c$ the constant for each Arrhenius term. For simplicity, we assume the transition rates to obey Arrhenius equations only depending on the activation energy. In Figure \ref{fig:sup.fig2} we found a qualitative solution to the model for a given set of parameters and it does not imply that it is the unique set of parameters that models the experimental data.

\begin{figure}
    \centering
    \includegraphics[width=1\textwidth]{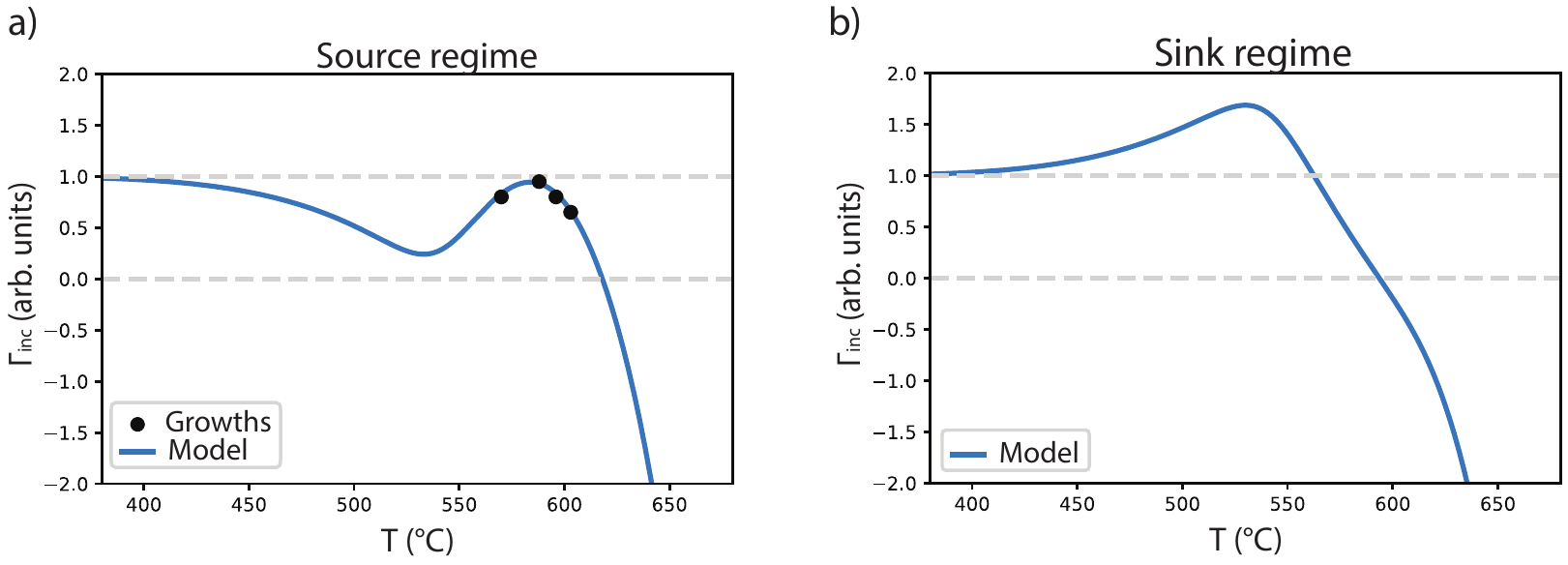}
    \vspace{-10 pt}
    \caption{Full T range of the source and sink NW growth modes, modelled using the growth equation eq. \ref{eq:model_Fig2d} from S1. (a) Modelled incorporation for source regime. (b) Modelled incorporation for sink regime.}
    \label{fig:sup.fig3}
\end{figure}

\section{S2. Longitudinal source effect and facet roughness} \label{S4}

\begin{figure}
    \centering
    \includegraphics[width=0.55\linewidth]{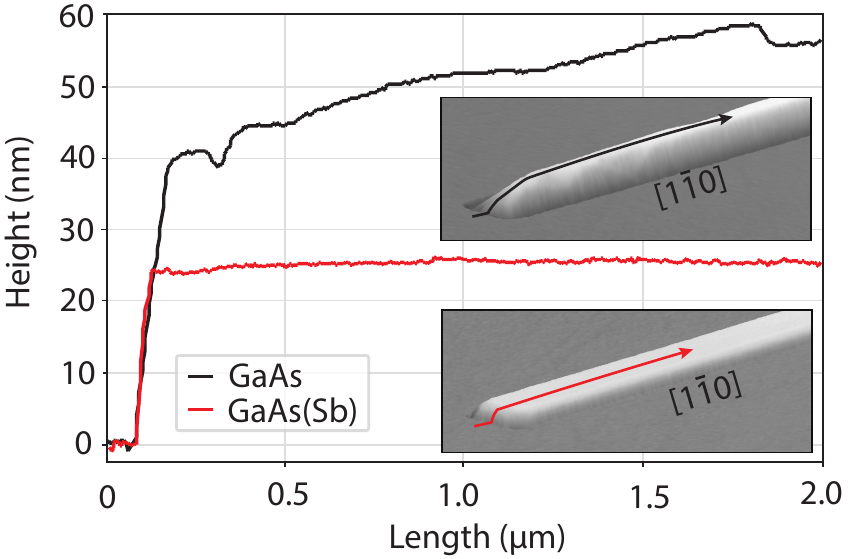}
    \caption{AFM longitudinal height profiles of GaAs and GaAs(Sb) 4 \textmu m long NWs along [1$\bar{1}$0]. The linescans show the effect in incorporation of the source effect near the NW ends. The GaAs incorporation steadily increases from the end until the half point of the NW.}
    \label{fig:sup.fig5}
\end{figure}

The volume measurements to calculate NW incorporation were performed in the middle part of the NW. Figure \ref{fig:sup.fig5} shows AFM height linescans of GaAs and GaAs(Sb) NWs inside arrays and the effect on the incorporation when approaching the end part. The reduction in incorporation is explained by the increase of available mask surrounding the NW at the end part, where adatoms can diffuse to and desorb $\Gamma_{a_mv}$.

\begin{figure}
    \centering
    \hspace*{-1cm}
    \includegraphics[width=1.1\linewidth]{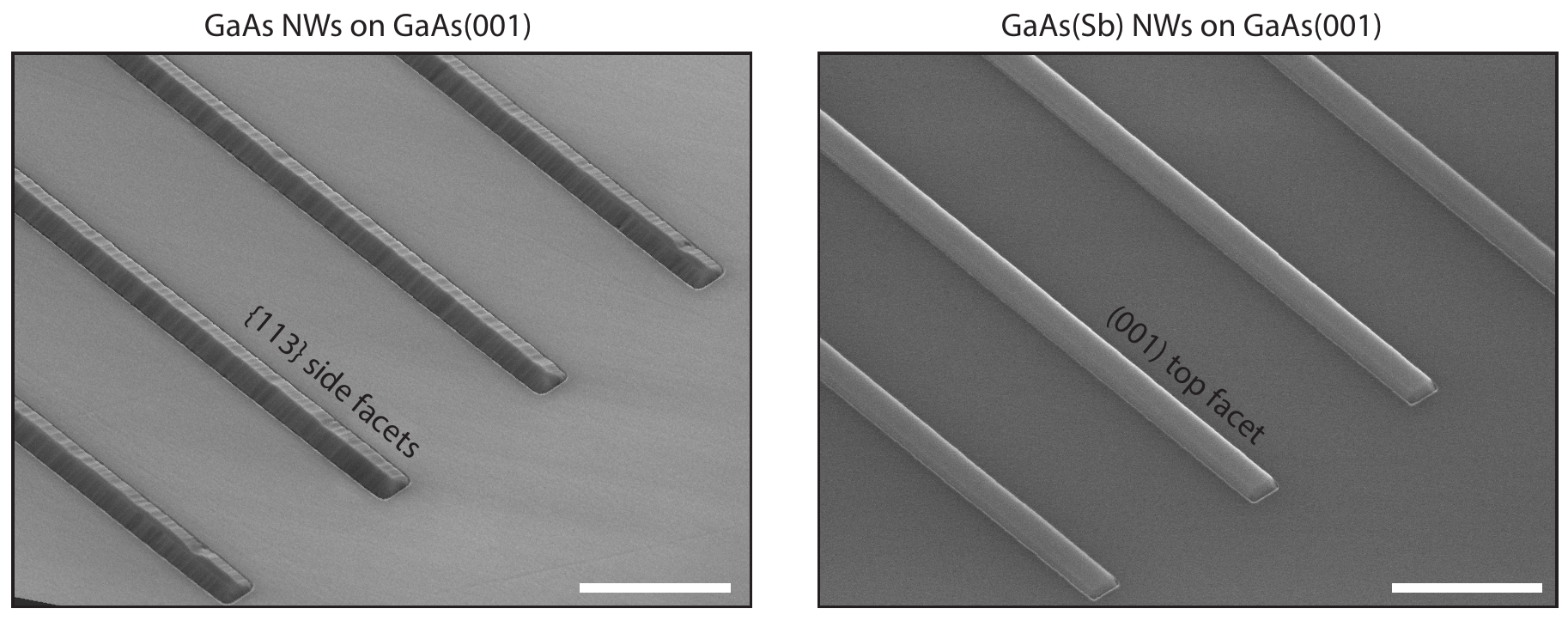}
    \caption{SEM images of pure GaAs and GaAs(Sb) NWs oriented along $[1\bar{1}0]$ crystallographic direction. The roughness of \{113\} side facets on GaAs is considerably higher than the top (001) for the GaAs(Sb) case. Scale bars are 1 \textmu m.}
    \label{fig:sup.fig8}
\end{figure}

Pure GaAs and GaAs(Sb) SAG NWs grown on GaAs(001) have different faceting with dominant \{113\} and (001), respectively. Figure \ref{fig:sup.fig8} shows SEM images of GaAs and GaAs(Sb) NW arrays (samples 2 and 1, check Supplementary S4 for details). Pure GaAs grows rough side facets whereas by adding Sb surfactant the faceting changes to a smooth (001), as shown as well in the AFM image in Figure \ref{fig:sup.fig5}.

\section{S3. Pregrowth fabrication and characterization} \label{S5}

\begin{figure}
    \centering
    \includegraphics[width=0.7\linewidth]{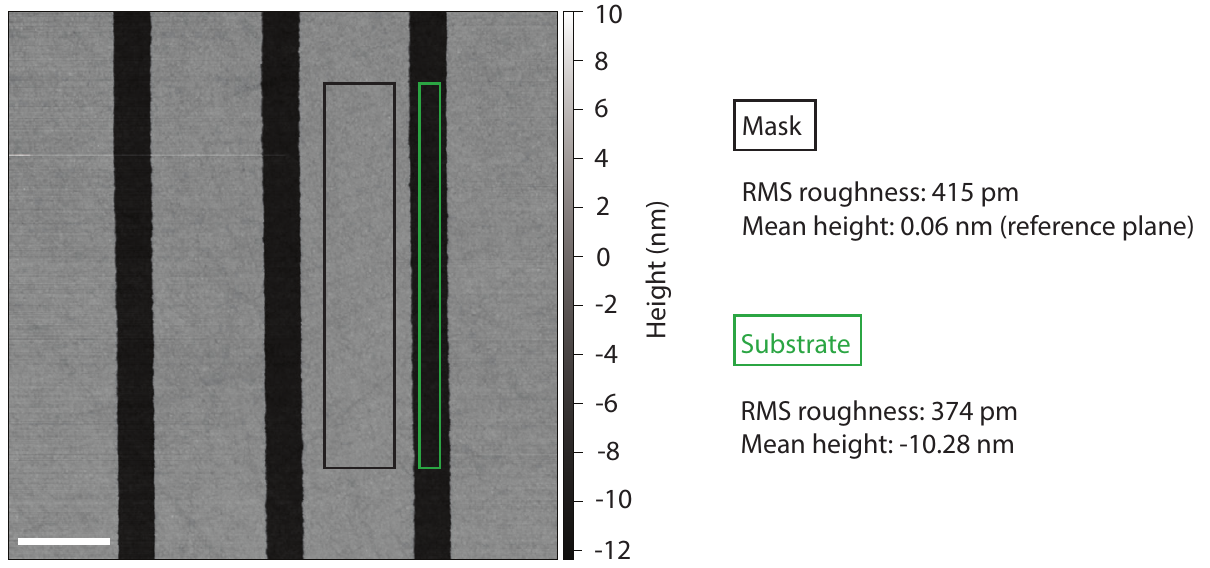}
    \caption{AFM image of the mask (black rectangle) and substrate (green rectangle) before MBE growth. Scale bar is 500 nm.}
    \label{fig:sup.fig6}
\end{figure}

AFM characterization is performed on the samples before growth. It is expected that different mask roughness affects the adatom kinetics on the mask during growth, specially the adatom diffusion length. The 10 nm height SiO$_2$ masks used in this study were deposited by Plasma Enhanced Chemical Vapor Deposition, as stated in references [12] and [33]. Figure \ref{fig:sup.fig6} shows the mean height of the mask and substrate and their roughness of a substrate before growth. Calculations of RMS roughness and mean height were performed with the AFM analysis software Gwyddion.
\\
\\

\section{S4. MBE growth recipes} \label{S2}

The following lists describe the growth details of the samples discussed in the main text: 

\begin{itemize}
    \item Figure 2a-c are growth recipes 1 (GaAs(Sb)) and 2 (GaAs).
    \item Figure 2d is a series of samples with common growth recipe 1, grown at the different temperatures.
    \item Figure 2e is growth recipe 1. These NWs were located at the edge of the growth substrate, unlike the rest of NWs in this work that are located in the central part. This fact had an impact on the local T of the substrate at the place of growth of Figure 2e NWs, being effectively smaller than the recorded from the MBE (603 °C) and showing some parasitic growth next to the NWs. It is expected that these NWs grown at the edge of the substrate show comparable source behavior than the ones in the center.
    \item Figure 5a is growth recipe 4. 
    \item Figure 5b is growth recipe 3.
\\

Growths:

    \begin{enumerate}
        \item GaAs(Sb): Growth T is 603 °C, As/Ga ratio 9, Sb/Ga ratio 3, Ga growth rate 0.1 ML/s, growth time 1800 s. SiOx mask (10 nm), grown on GaAs(001).
        \item GaAs: Growth T is 597 °C, As/Ga ratio 9, Ga growth rate 0.1 ML/s, growth time 1800 s. SiOx mask (10 nm), grown on GaAs(001).
        \item GaAs(Sb)-InGaAs-InAs: 
            \begin{enumerate}
                \item Growth T of GaAs(Sb) is 602 °C, As/Ga ratio 10.1, Sb/Ga ratio 3, Ga growth rate 0.1 ML/s, growth time 1800 s.
                 \item Growth T of InGaAs is 533 °C, As/Ga ratio 12, Ga growth rate 0.01, In growth rate 0.09 ML/s, growth time 3600 s.
                 \item Growth T of InAs is 523 °C, As/In ratio 9, In growth rate 0.09 ML/s.
                 AlOx (2nm) - SiOx (10nm) mask, grown on GaAs(001).
            \end{enumerate}
         \item In$_{0.53}$Ga$_{0.47}$As: Growth T is 508 °C, In growth rate 0.023 ML/s, Ga growth rate 0.01 ML/s, growth time 7200 s.
         
    \end{enumerate}
\end{itemize}

\section{S5. Characterization techniques} \label{S3}

AFM post-growth characterization of the samples was perfomed in an Bruker Icon AFM using tapping PeakForce Quantitative Nanoscale Mechanical Characterization mode. The probes used were Bruker Scanasyst-Air.

Annular Dark Field Scanning Transmission Electron Micrographs were acquired on a 300 kV acceleration voltage FEI Titan using the High Angle Annular detector. Lamella sample preparation of the NWs cross section was performed on a FEI Helios 600 dual beam system.

Figure S4 shows the raw ADF-STEM image presented in Figure 5 in the main text, with the material contrast used to determine the height of the different layers.

\begin{figure}[h]
    \centering
    \includegraphics[width=0.4\linewidth]{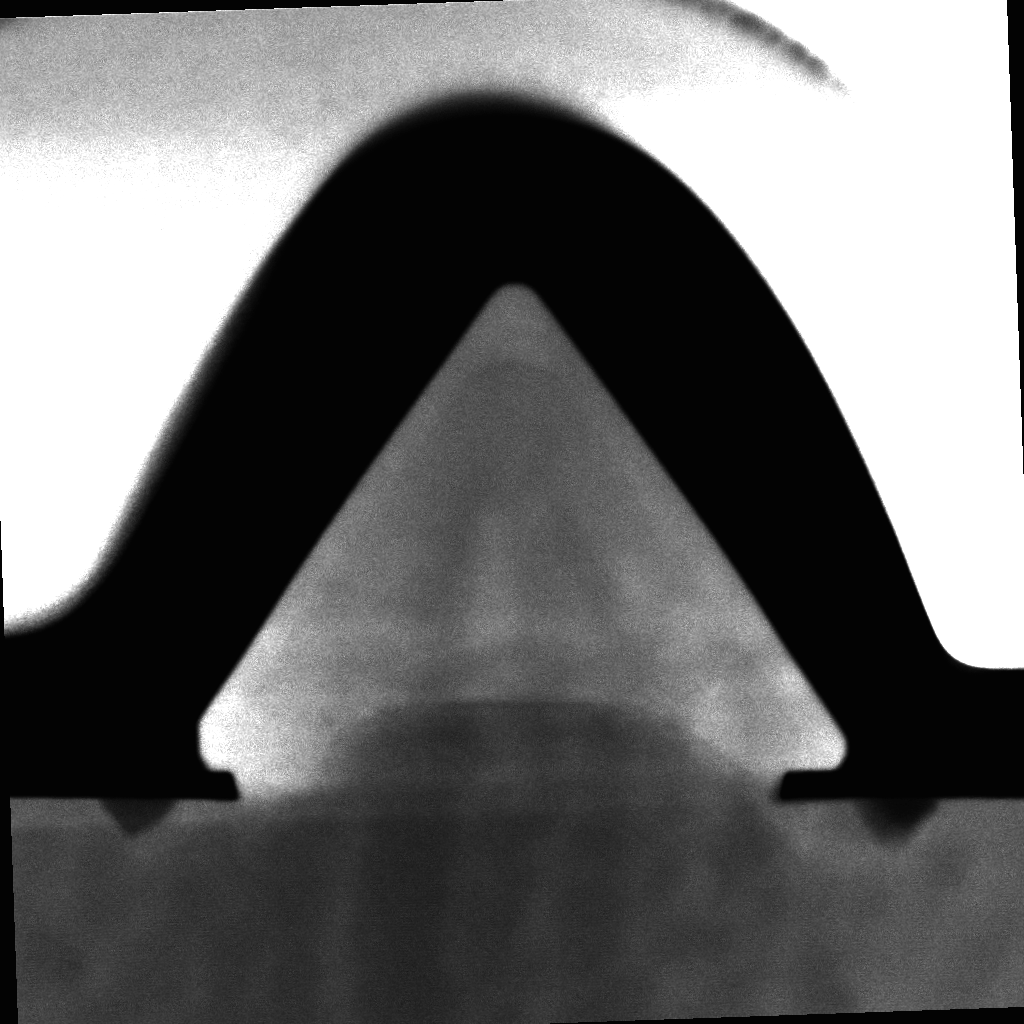}
    \caption{ADF-STEM raw image from Figure 5b. The dark and bright layers on top of the NW cross section correspond to C and Pt depositions deposited for protection purposes. Frame width is 400 nm.}
    \label{fig:sup.fig4}
\end{figure}

\section{S6. Selectivity between arrays of nanowires} \label{S6}

\begin{figure}[h]
    \centering
    \includegraphics[width=1\linewidth]{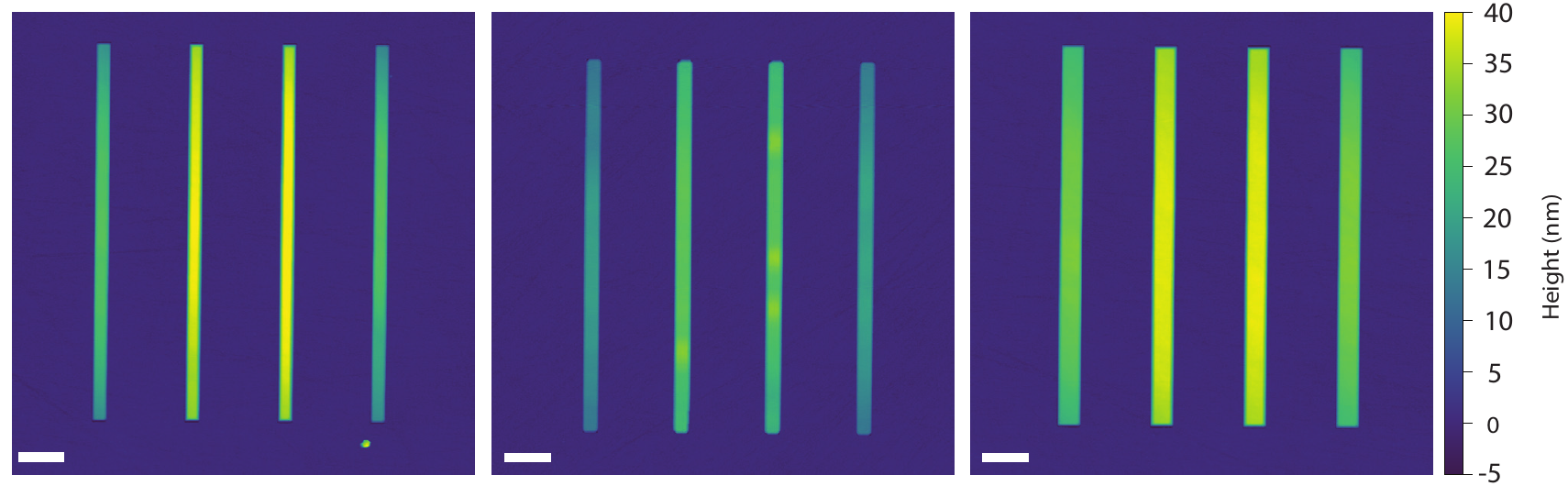}
    \caption{AFM images of arrays of GaAs(Sb) NWs with varying $w$. These arrays belong to the same sample from the single NWs in Figure 2e and are located at the central region of the substrate, showing the absence of crystallites between NWs. NWs are oriented along the $[1\bar{1}0]$ crystallographic direction and scale bars are 500 nm.}
    \label{fig:sup.fig7}
\end{figure}

The NW volume and selectivity measurements for the samples detailed in Section \nameref{S2} are performed in the central region of each growth substrate. Figure \ref{fig:sup.fig7} shows 3 AFM images of GaAs(Sb) NW arrays with different width grown at the central part of the substrate, from sample 1 (see Section \nameref{S2} for details). The selectivity is good in between the NWs, as expected from the sample growth T.

\section{S7. Crystal desorption measurements} \label{S7}

\begin{figure}
    \centering
    \includegraphics[width=0.7\linewidth]{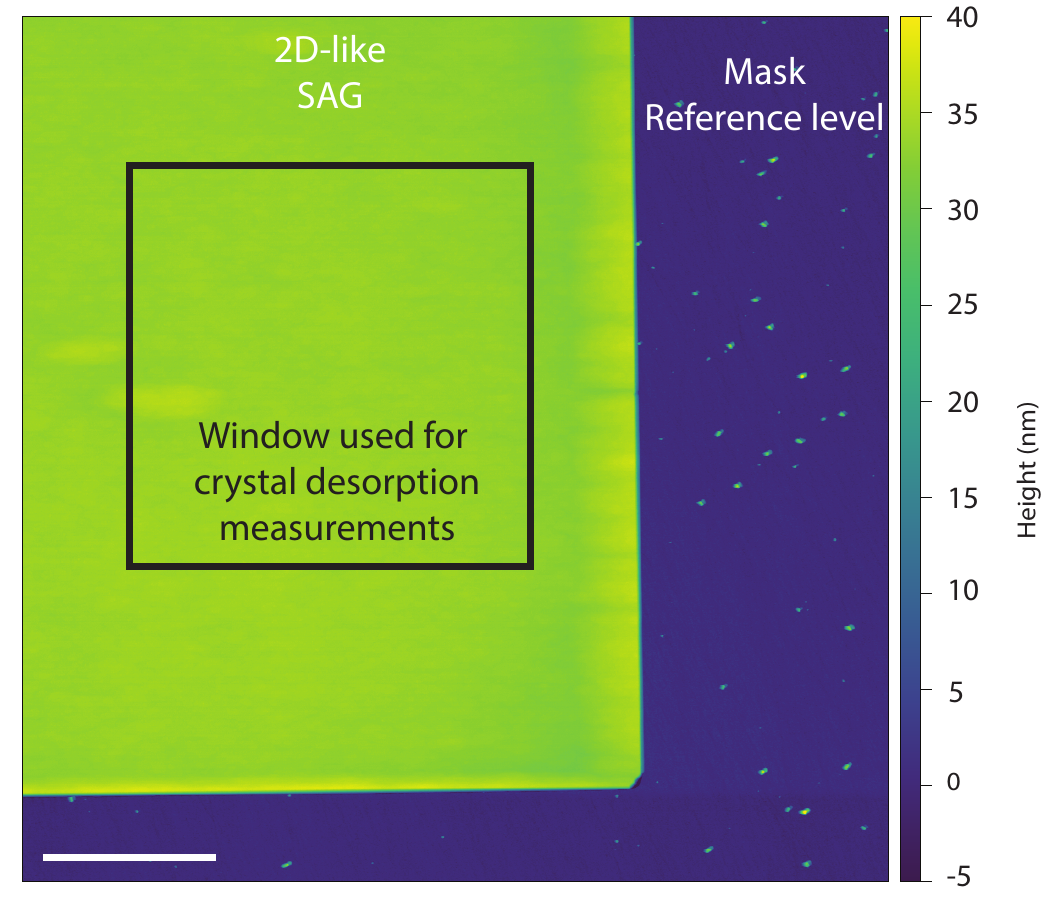}
    \caption{AFM image of a big mask opening used to measure the adatom desorption from the crystal. The image is leveled with the mask surface and the volume is measured in the center of the structure. The sample is GaAs(Sb) grown at 570 °C. Scale bar is 2 \textmu m.}
    \label{fig:sup.fig9}
\end{figure}

Crystal adatom desorption data from Figure 2d is measured in big mask openings grown in the center of the substrate. Figure \ref{fig:sup.fig9} is an AFM image of the corner of a 10x10 \textmu m mask opening from a GaAs(Sb) sample grown at 570 °C. Using the mask as a reference height, the volume of a 5x5 \textmu m square is measured and compared with the equivalent volume from a planar growth sample used for growth rate calibrations.

\newpage